\documentclass[
    ,final            
  ]
  {aipproc}

\usepackage{amsmath}

\layoutstyle{6x9}

\begin{document}

\title{Rapidity dependence of HBT correlation radii in non-boost invariant models}

\classification{25.75.-q}
\keywords      {HBT, correlations, rapidity dependence}

\author{Thorsten Renk}{
  address={Department of Physics, Duke University, PO Box 90305,  Durham, NC 27708 , USA}
}

\begin{abstract}
Hanbury-Brown Twiss (HBT) correlation measurements provide valuable information about the
phase space distribution of matter in ultrarelativistic heavy-ion collisions. The rapidity
dependence of HBT radii arises from a nontrivial interplay between longitudinal and transverse expansion
and the time dependence of the freeze-out pattern. For a non-accelerating
longitudinal expansion the dependence primarily arises from the amount of radiating matter
per unit rapidity $dN/d\eta$, but for a scenario with strong longitudinal acceleration  
additional complications occur. In this paper I explore schematically
what type of dependence can be expected for RHIC conditions under different model assumptions
for the dynamics of spacetime expansion and freeze-out.
\end{abstract}

\maketitle


\section{Introduction}

While the final state of a ultrareativistic heavy-ion collision is to a large degree 
reflected in the measured distribution of hadrons there are different possible evolutions 
leading to this final state which leave a more subtle imprint on observables such as transverse 
mass $m_T = \sqrt{m^2 +p_T^2}$ spectra and two particle correlations.

The two extreme assumptions are the boost-invariant model proposed by Bjorken \cite{Bjorken}
in which the incoming nuclei are nearly transparent and the Landau scenario \cite{Landau} in which the incoming 
nuclei are completely stopped and the resulting extremely hot and dense system is assumed 
to undergo strongly accelerated expansion. In a Bjorken expansion matter moves on
free streaming trajectories from a source with negligible longitudinal extension (the initial long. nuclear overlap), 
leading to rapidity $\eta = \frac{1}{2}\frac{p_0 + p_z}{p_0-p_z}$ being equal
to spacetime rapidity $\eta_s =  \frac{1}{2}\frac{t + z}{t-z}$. In this scenario a plateau in the 
multiplicity per unit rapidity $dN/d\eta$ is expected around $\eta=0$.
In contrast, it has been argued that $dN/d\eta$ at RHIC can reasonably well be described with 
Landau hydrodynamcis \cite{Steinberg}. In this scenario, $\eta_s \neq \eta$, no plateau is expected 
and $dN/d\eta$ changes strongly in time as the system undergoes longitudinal acceleration.

Most 3d hydrodynamic calculations start with initial conditions close to Bjorken-like free streaming
(albeit without a plateau in rapidity) and consequently do not alter $dNd\eta$ significantly during the evolution
\cite{3dh}. In contrast, in \cite{RHIC} I have suggested that a significant longitudinal 
acceleration seems to be necessary if a simultaneous description of transverse mass spectra and HBT
correlation radii is required.

In a longitudinal free-flow scenario (with $dN/d\eta$ approximately unchanged), 
the effect dominating the rapidity dependence of Hanbury-Brown-Twiss (HBT) radii and 
$m_T$-spectra is the shape of $dN/d\eta$ ---
the amount of thermalized matter at given $\eta$ governs the amount of expansion and  flow 
that can be reached before decoupling.

In contrast, in a scenario in which $dN/d\eta$ changes strongly over time, radiation from the expanding
system will only contribute to the yield at forward rapidities in later evolution stages
as matter accelerates outward. Thus, e.g. the
duration of the presence of thermalized matter at forward rapidities will be potentially be 
smaller than at midrapidity quite independent
from variations in $dN/d\eta$. However, spectra and HBT radii will only reflect this if emission of particles
during the evolution time is a significant fraction of the total particle emission inlcuding final
breakup. It is the aim of this paper to study to what degree the rapidity dependence of HBT correlation radii 
is capable of reflecting the evolution history of the system.

\section{Model framework}

I use the model described in \cite{RHIC, Synopsis} for the analysis.
Its main assumption is that an equilibrated system is formed
a short time $\tau_0$ after the onset of the collision. This 
thermal fireball subsequently expands isentropically until the mean free path of particles exceeds
the dimensions of the system and particles decouple at a timescale $\tau_f$.

For the entropy density at a
given proper time we make the ansatz 
\begin{equation}
s(\tau, \eta_s, r) = N R(r,\tau) \cdot H(\eta_s, \tau)
\end{equation}
with $\tau $ the proper time as measured in a frame co-moving
with a given volume element  and $R(r, \tau), H(\eta_s, \tau)$ two functions describing the shape of the distribution
and $N$ a normalization factor.
Woods-Saxon distributions
\begin{equation}
R(r, \tau) = 1/\left(1 + \exp\left[\frac{r - R_c(\tau)}{d_{\text{ws}}}\right]\right), \quad
H(\eta_s, \tau) = 1/\left(1 + \exp\left[\frac{\eta_s - H_c(\tau)}{\eta_{\text{ws}}}\right]\right).
\end{equation}
are used to describe the shapes for given $\tau$. Thus, the ingredients of the model are the 
skin thickness parameters $d_{\text{ws}}$ and $\eta_{\text{ws}}$
and the para\-me\-tri\-zations of the expansion of the spatial scales $R_c(\tau), H_c(\tau)$ 
as a function of $\tau$.  For the radial expansion assuming constant acceleration one finds
$R_c(\tau) = R_0 + \frac{a_\perp}{2} \tau^2$. $H_c(\tau)$ is obtained
by integrating forward in $\tau$ a trajectory originating from the collision center which is characterized
by a rapidity  $\eta_c(\tau) = \eta_0 + a_\eta \tau$
with $\eta_c = \text{atanh } v_z^c$ where $v_z^c$ is  the longitudinal velocity of that 
trajectory. Since the relation between proper time as measured in the co-moving frame 
and lab time is determined by the rapidity at a given time, the resulting integral is
in general non-trivial and solved numerically (see \cite{Synopsis} for details).
$R_0$
is determined in overlap calculations using Glauber theory, the initial size of the rapidity interval occupied
by the fireball matter.  The set of model parameters then includes initial rapidity $\eta_0$
transverse velocity  $v_\perp^f = a_\perp \tau_f$ and
rapidity at decoupling proper time $\eta^f = \eta_0 + a_\eta \tau_f$.
Thus, specifying $\eta_0, \eta_f, v_\perp^f$ and $\tau_f$ sets the scales of the spacetime
evolution and $d_{\text{ws}}$ and $\eta_{\text{ws}}$ specify the detailed distribution of entropy density.

For transverse flow a linear relation between radius $r$ and
transverse rapidity $\rho = \text{atanh } v_\perp(\tau) =  r/R_c(\tau) \cdot \rho_c(\tau)$ is assumed
with $\rho_c(\tau) = \text{atanh } a_\perp \tau$.
The model allows for the possibility of accelerated longitudinal expansion
which in general implies $\eta \neq \eta_s$ \cite{Synopsis}.  This mismatch between
spacetime and momentum rapidity can be parametrized as
a local $\Delta \eta = \eta -\eta_s$ which is a function of $\tau$ and $\eta_s$.

With the help of a quasiparticle equation of state (EOS), one can find the local 
temperature $T(\eta_s, r, \tau)$ 
from the entropy density $s(\eta_s, r, \tau)$ and net baryon density $\rho_B(\eta_s,r,\tau)$.
Particle emission is calculated throughout the lifetime of the fireball
by selecting a freeze-out temperature $T_f$, finding the hypersurface characterized by
$T(\eta_s, r, \tau) = T_f$ and evaluating the Cooper-Frye
formula
\begin{equation}
E \frac{d^3N}{d^3p} =\frac{g}{(2\pi)^3} \int d\sigma_\mu p^\mu
\exp\left[\frac{p^\mu u_\mu - \mu_i}{T_f}\right] = d^4 x S(x,p)
\end{equation}
with $p^\mu$ the momentum of the emitted particle and $g$ its degeneracy factor.

HBT correlation radii are determined as averages of coordinates with the emission function \cite{HBTReport}

\begin{equation}
R_{side}^2({\bf K}) = \langle \tilde{y}^2 \rangle({\bf K}), \quad R_{out}^2({\bf K}) = \langle (\tilde{x}- \beta_\perp \tilde{t})^2 \rangle({\bf K}) \quad \text{and} \quad R_{long}^2({\bf K}) = \langle \tilde{z}^2 \rangle ({\bf K}) 
\end{equation}
where
\begin{displaymath}
\tilde{x}^{\mu}(K) = x^\mu - \langle {x}^\mu\rangle(K) \quad \text{with}
\quad \langle f \rangle(K) = \frac{\int d^4 x f(x) S(x,K)}{\int d^4x S(x,K)}.
\end{displaymath}

By choosing $\eta_0$ different from the measured width of $dN/d\eta$ it is possible to  introduce longitudinal
acceleration into the model. Note that the model description cannot be assumed to be valid in the target/projectile
fragmentation --- as one probes forward rapidities, thermal physics is less relevant, evident from the
fact that a large fraction of matter is below $T_F$ {\it ab initio}. Thus, the model prediction beyond $\eta =2$ 
should be taken with some caution.

\section{$R_{out}$ and the freeze-out pattern}

\begin{figure}
  \includegraphics[width=5.9cm]{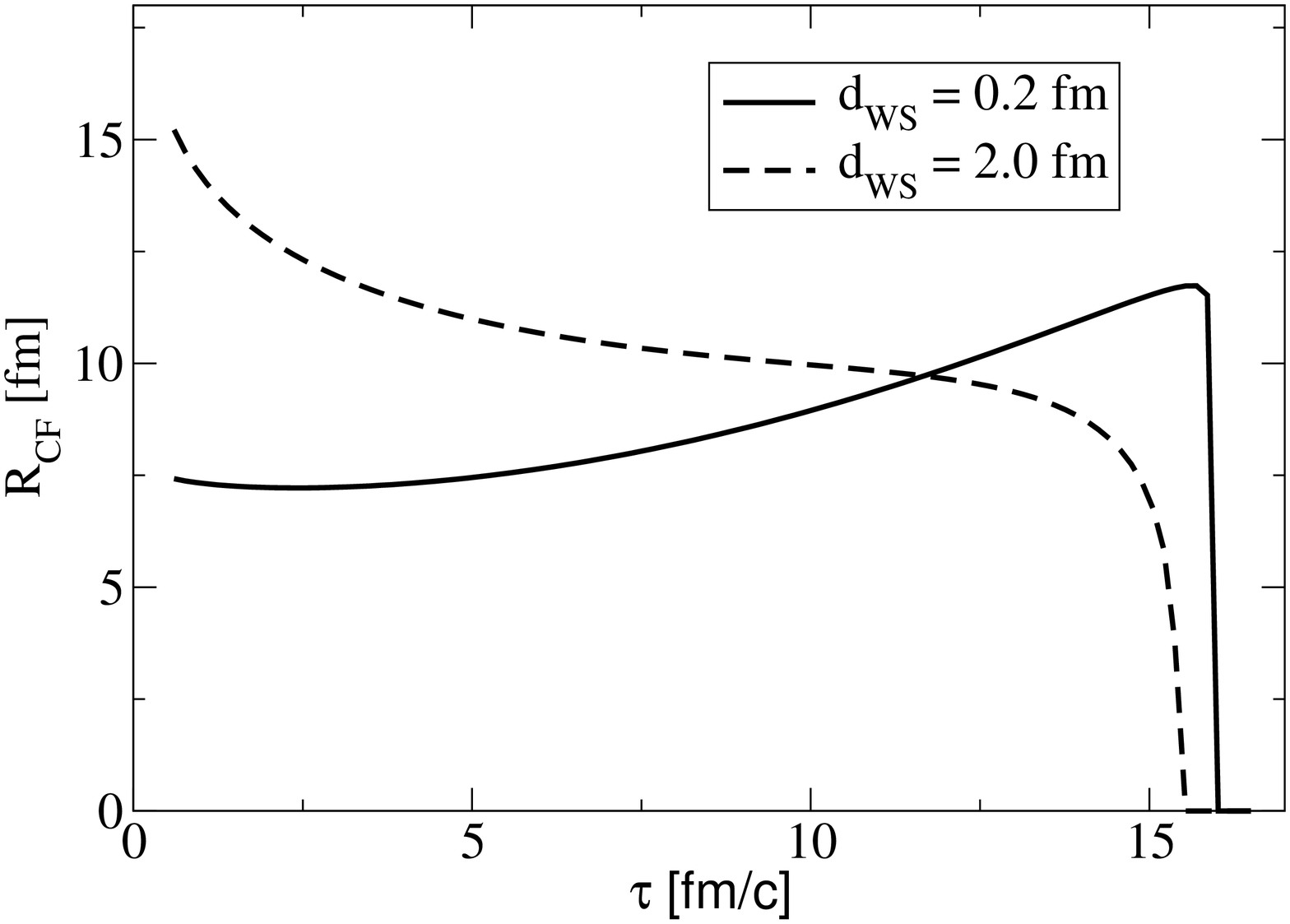}
  \includegraphics[width=6.0cm]{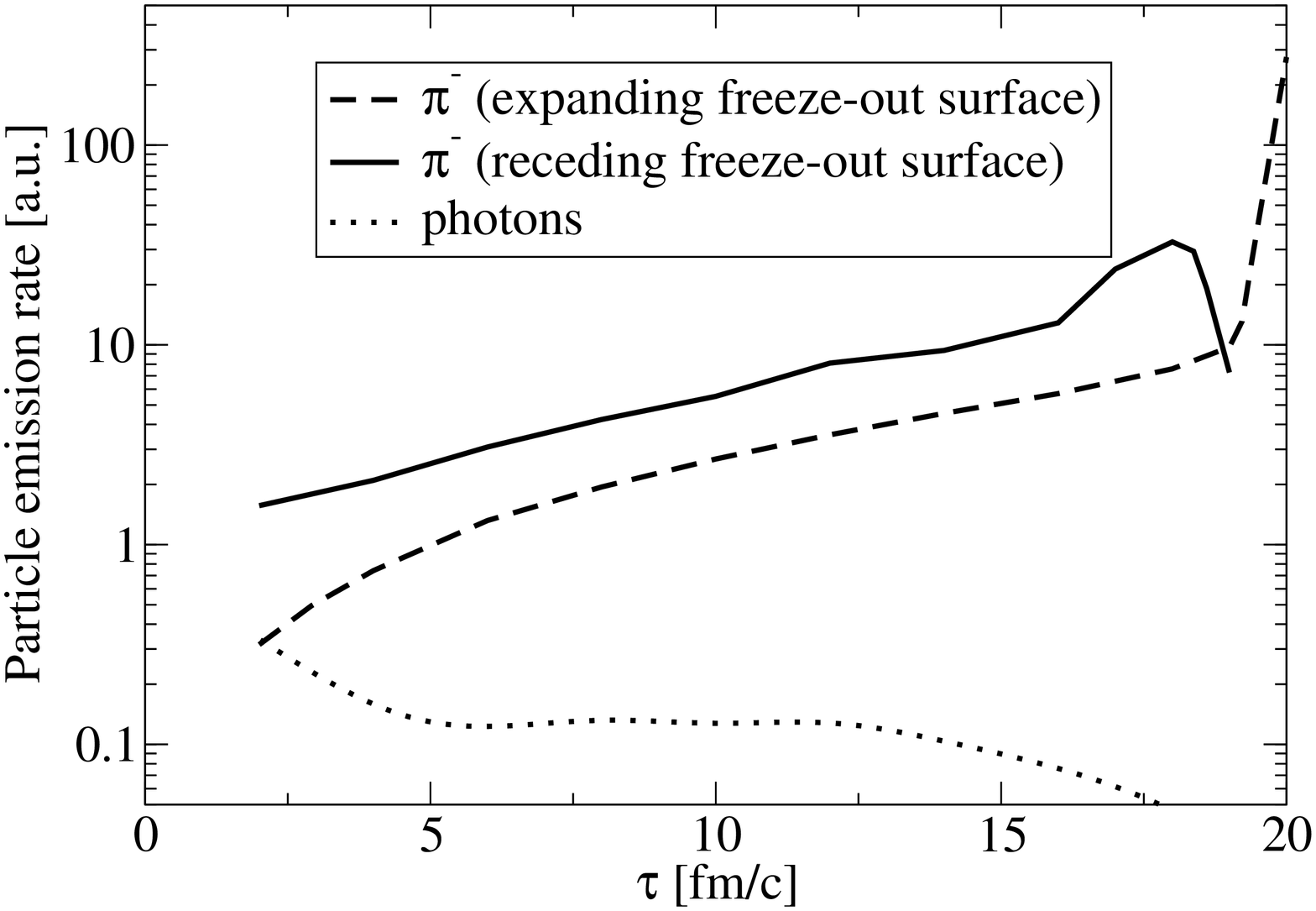}

  \caption{\label{F-fo}Left panel: Evolution of the Cooper-Frye surface radius $R_{CF}$ at $\eta=0$ 
for $T_F=110$ MeV as a function of $\tau$ for different values of radial skin thickness parameter. 
Right panel: Particle emission (in arbitrary units) 
into midrapidity as a function of $\tau$ for pions as a function of $R_{CF}(\tau)$ and for photons.}
\end{figure}

We expect that the spacetime evolution of the emission source is reflected in the $\eta$
dependence of HBT radii only if the system is not dominated by the final breakup. Since freeze-out
is determined dynamically, there is no apparent parameter to adjust amount of hadrons released by the
final breakup vs. the amount of particles continuously emitted before. From Fig.~\ref{F-fo} it is however evident
that selecting the skin thickness $d_{ws}$ in the radial distribution of entropy
density has a visible impact on the radial position $R_{CF}(\tau)$ of the Cooper-Frye surface.

$R_{CF}$ shrinks for a large skin thickness (almost a Gaussian) whereas it grows for a
small skin thickness (a box-like distribution).
Since particles must move faster than the Cooper-Frye surface in order to be released, an inward-moving surface
implies much more early emission than an outward moving one. This is confirmed in the right panel of Fig.~\ref{F-fo} where we 
present the number of particles being released as a function of $\tau$.
The width of the final peak in this figure is connected with the emission time, i.e. the difference in $R_{out}$ and
$R_{side}$. Note that the data seem to prefer the solution with expanding $R_{CF}$, i.e. a sharp surface, at least in
the last stages of the evolution. Shockwaves in the medium excited by hard parton energy loss as observed
in two particle correlations \cite{PHENIX, Machcones} could be a potential
mechanism for creating such a sharply defined surface.

\section{Results}

I investigate rapidity dependence in three different scenarios designed to show the essential effects to be expected:
1) a longitudinal free-flow expansion where $dN/d\eta$ remains approximately constant during the evolution
2) a longitudinally accelerated scenario where the width of $dN/d\eta$ changes from 1.7 to 3.8 during the
evolution and continuous particle emission occurs and 
3) a longitudinally accelerated expansion with the same parameters as above but a sudden breakup in the end (note that
case 1) is also characterized by a sudden breakup).
In all three cases single particle $m_T$ spectra describe the data at midrapidity and the measured $dN/d\eta$ is reproduced,
though only 3) describes the measured HBT correlation radii at midrapidity correctly.

\begin{figure}
  \includegraphics[width=5cm]{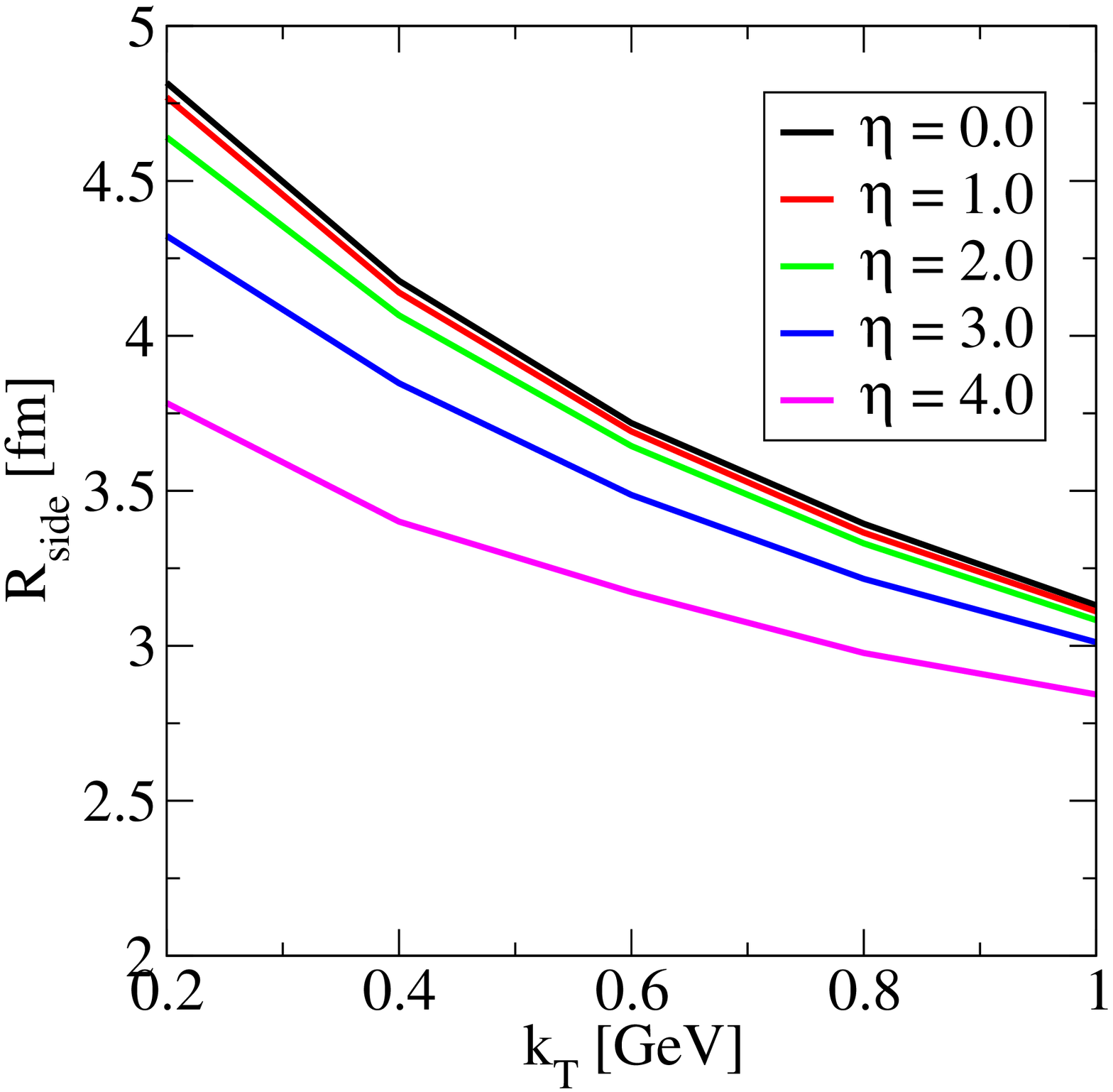}
  \includegraphics[width=5cm]{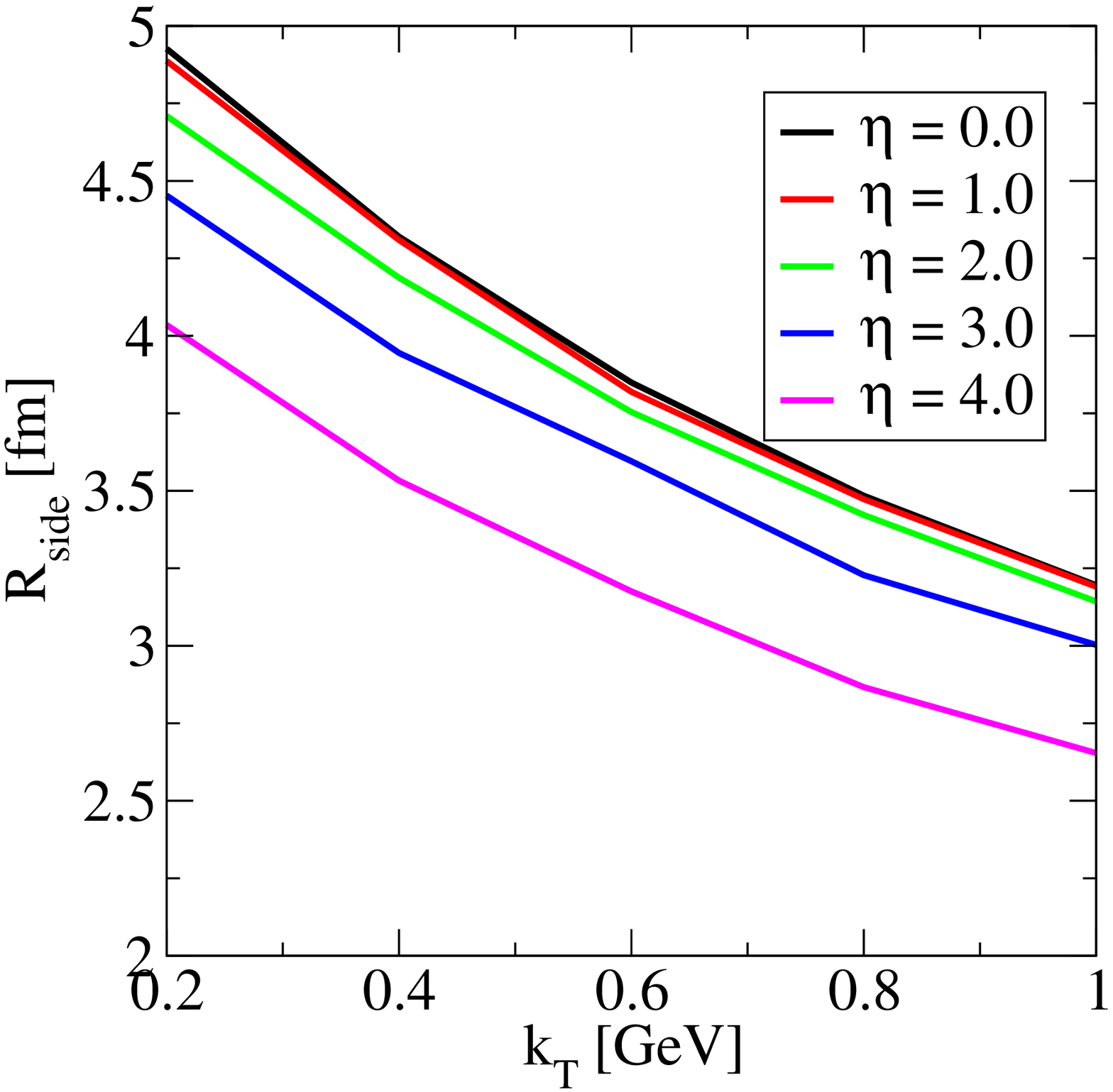}
  \includegraphics[width=4.8cm]{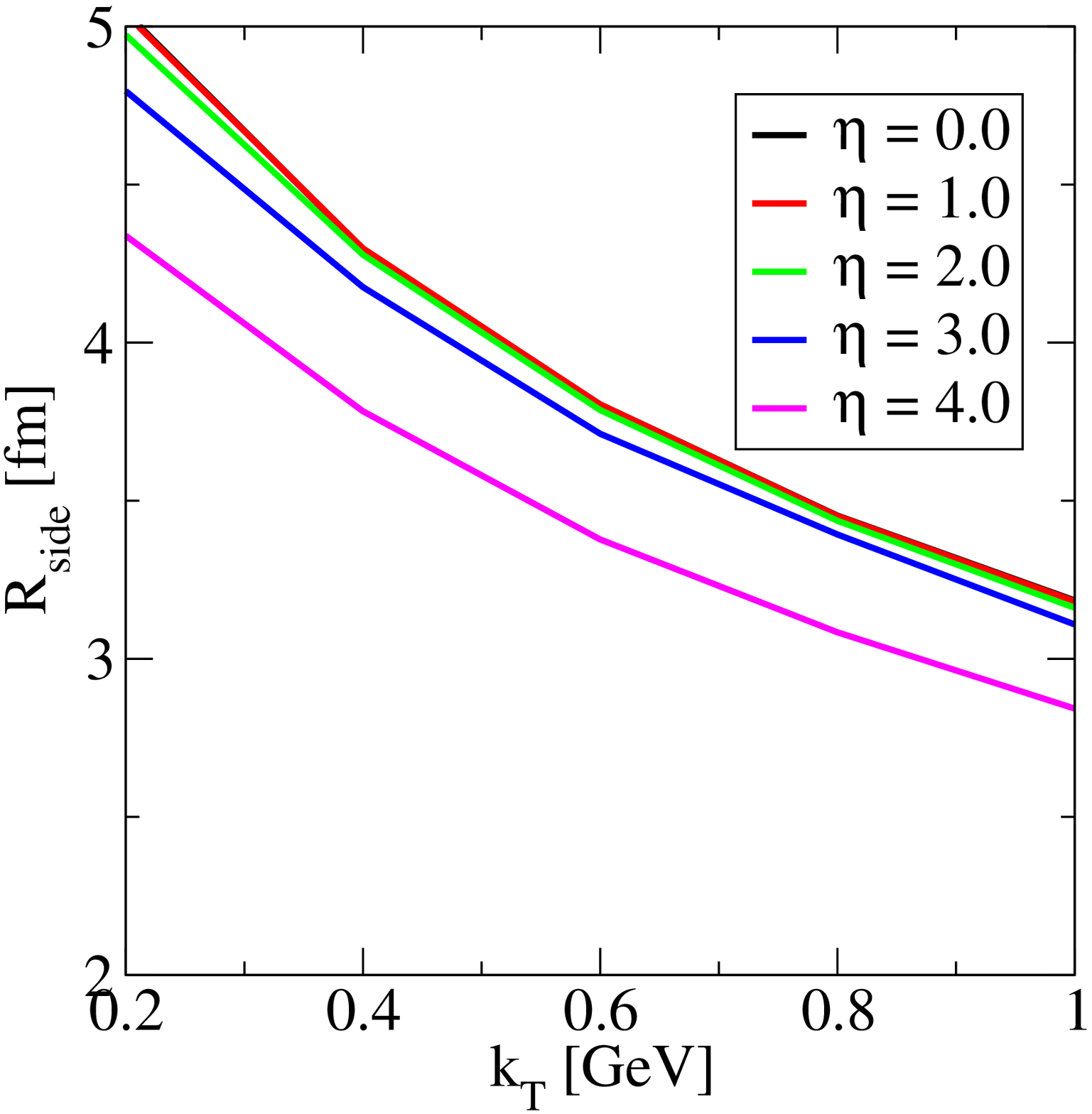}

  \caption{\label{F-side}$R_{side}$ as a function of pair momentum $k_T$ for different rapidities $\eta$ in 
an approximate boost invariant expansion (left), a longitudinally accelerating scenario with continuous particle
emission (middle) and a longitudinally accelerating scenario with sudden breakup (right).}
\end{figure}

Figure~\ref{F-side} shows $R_{side}$ in the three different scenarios as a function of
pair transverse momentum for different rapidities $\eta$.


The differences between the three scenarios is generally small, however the geometrical radius 
in the free-flow expansion is systematically smaller than in the accelerated cases. This is due
to the fact that breakup occurs roughly at a fixed volume (determined by the total entropy), hence the stronger
longitudinal expansion in the Bjorken case (which expands with $\eta_0$ at all times) implies less transverse
radius expansion and flow.

\begin{figure}
  \includegraphics[width=5cm]{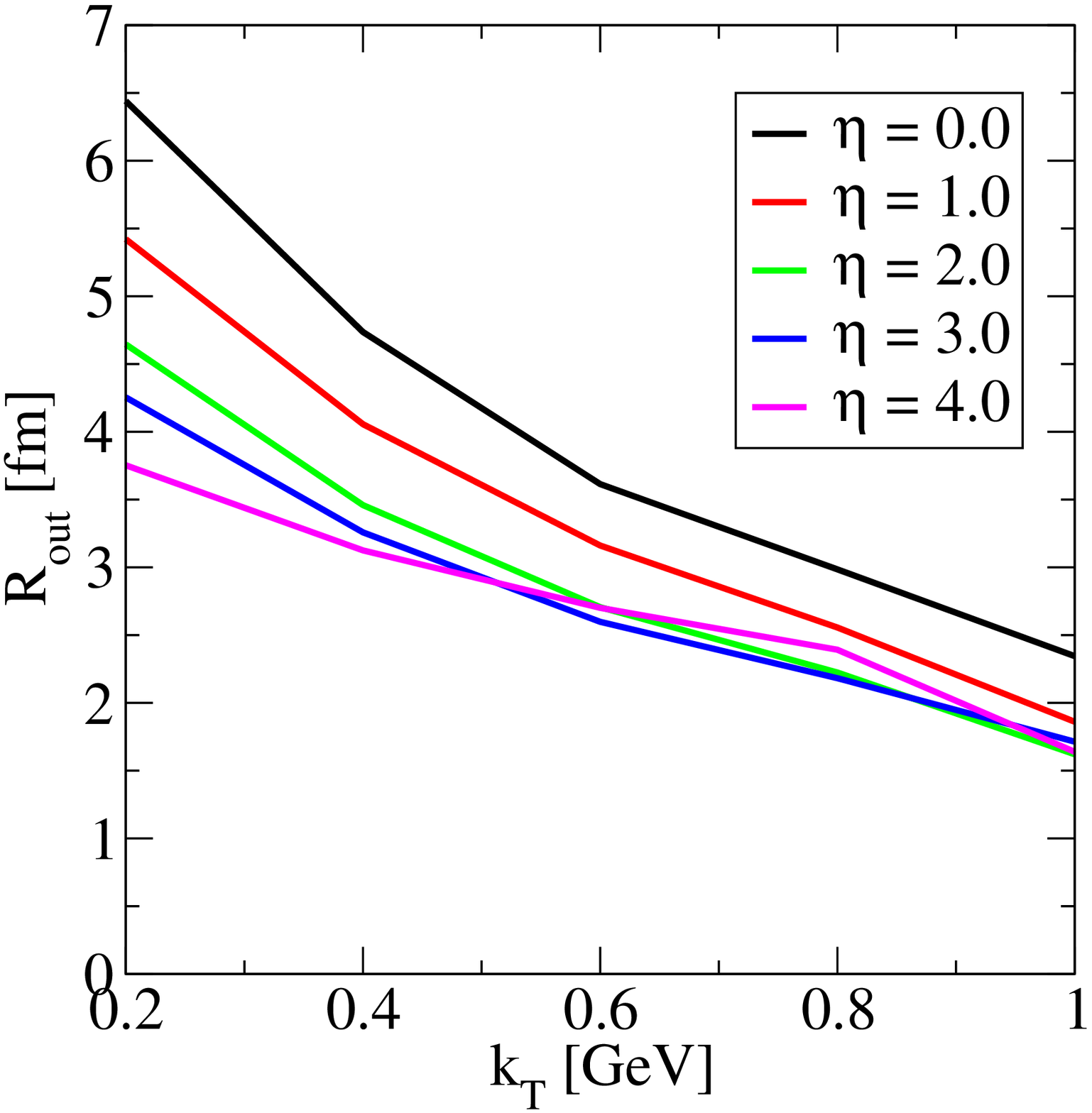}
  \includegraphics[width=5cm]{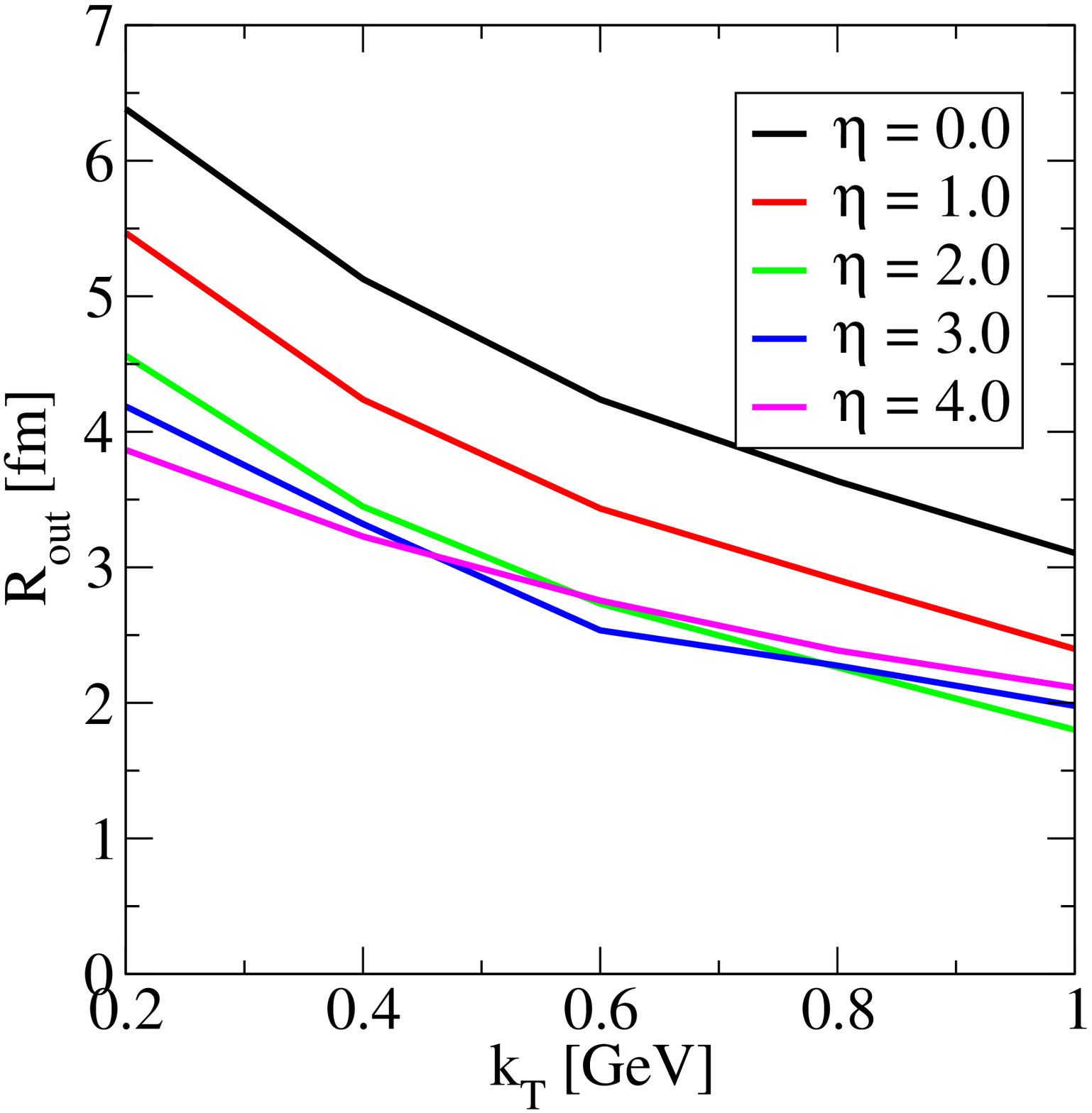}
  \includegraphics[width=5.0cm]{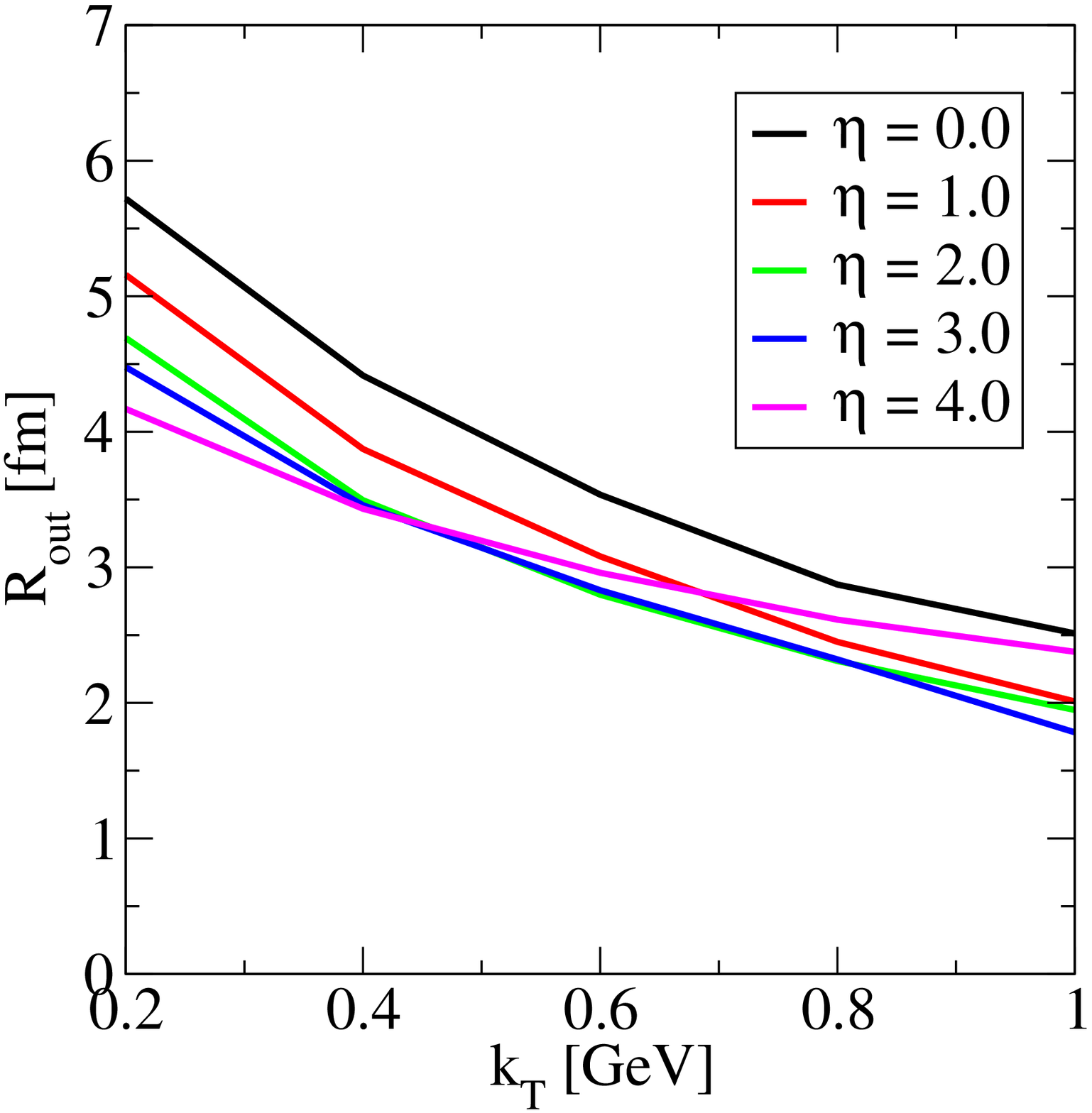}

  \caption{\label{F-out}$R_{out}$ as a function of pair momentum $k_T$ for different rapidities $\eta$ in 
an approximate boost invariant expansion (left), a longitudinally accelerating scenario with continuous particle
emission (middle) and a longitudinally accelerating scenario with sudden breakup (right).}
\end{figure}

Fig.~\ref{F-out} compares $R_{out}$ for the three different scenarios. The most striking effect 
is the general increase in the free flow and the continuous emission scenario as compared to the sudden
breakup one. The reason can easily be identified: With an expanding Cooper-Frye surface and a large fraction
of energy going into transverse expansion, the sudden breakup scenario has the strongest positive $x-t$ correlation
and the least release of particles prior to final breakup.
While sudden breakup also occurs in the approximate Bjorken solution, the stronger longitudinal expansion
leaves less energy for transverse expansion, significantly lessening this correlation. Finally, since the Cooper-Frye surface
is moving inward in the continuous emission scenario, there is no such correlation and $R_{out}$ significantly
exceeds $R_{side}$.

\begin{figure}
  \includegraphics[width=5cm]{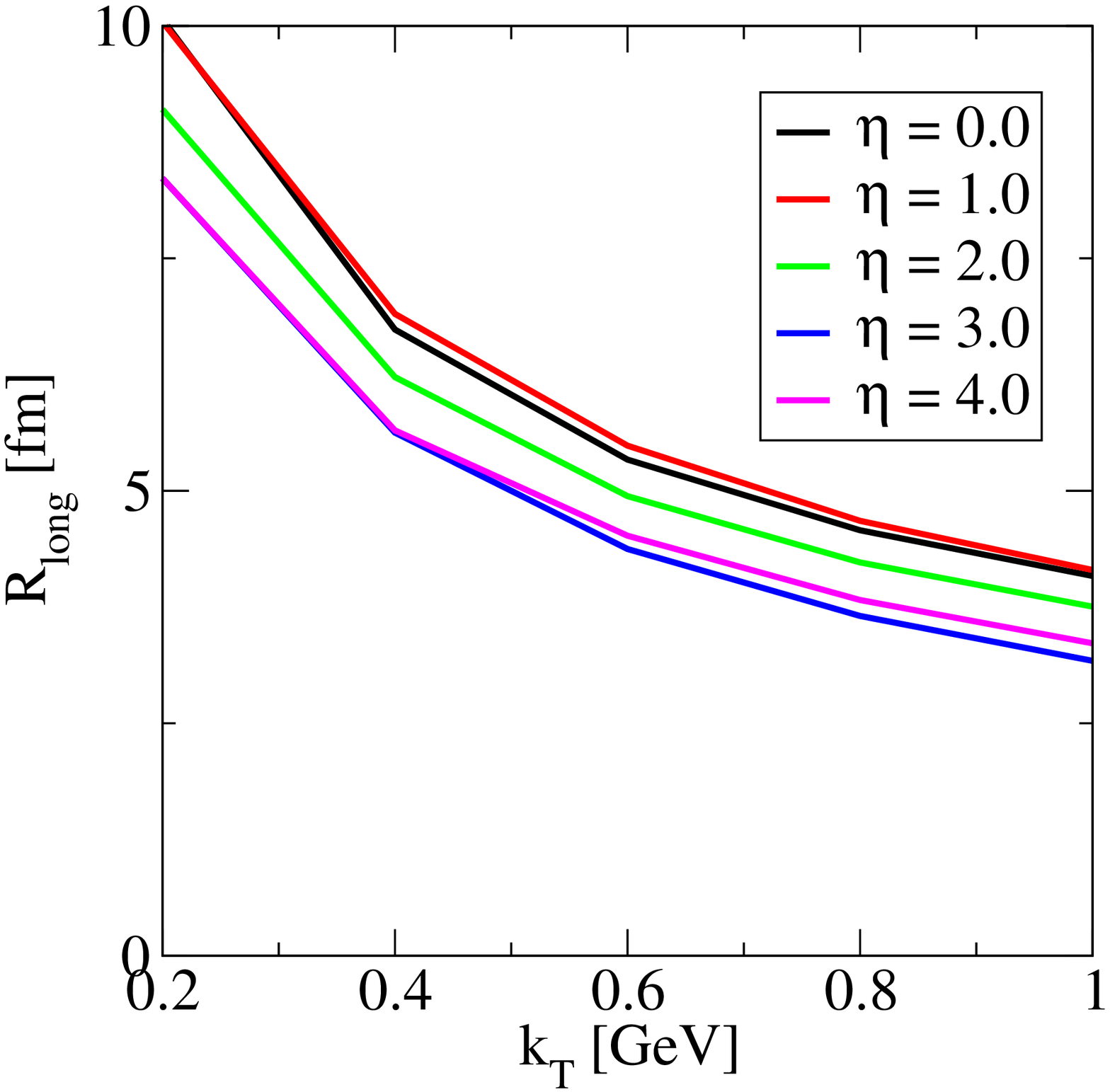}
  \includegraphics[width=5cm]{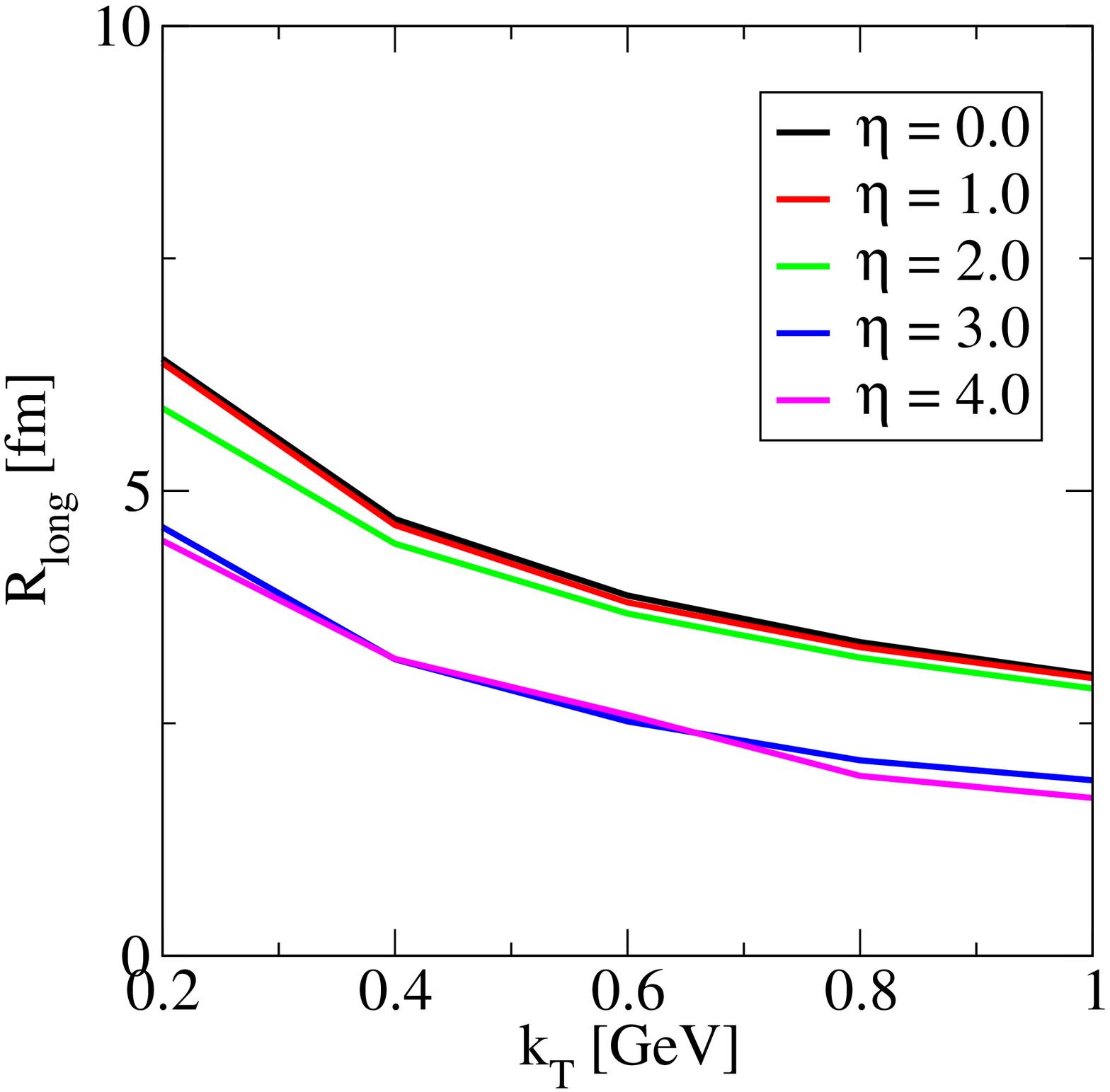}
  \includegraphics[width=5.0cm]{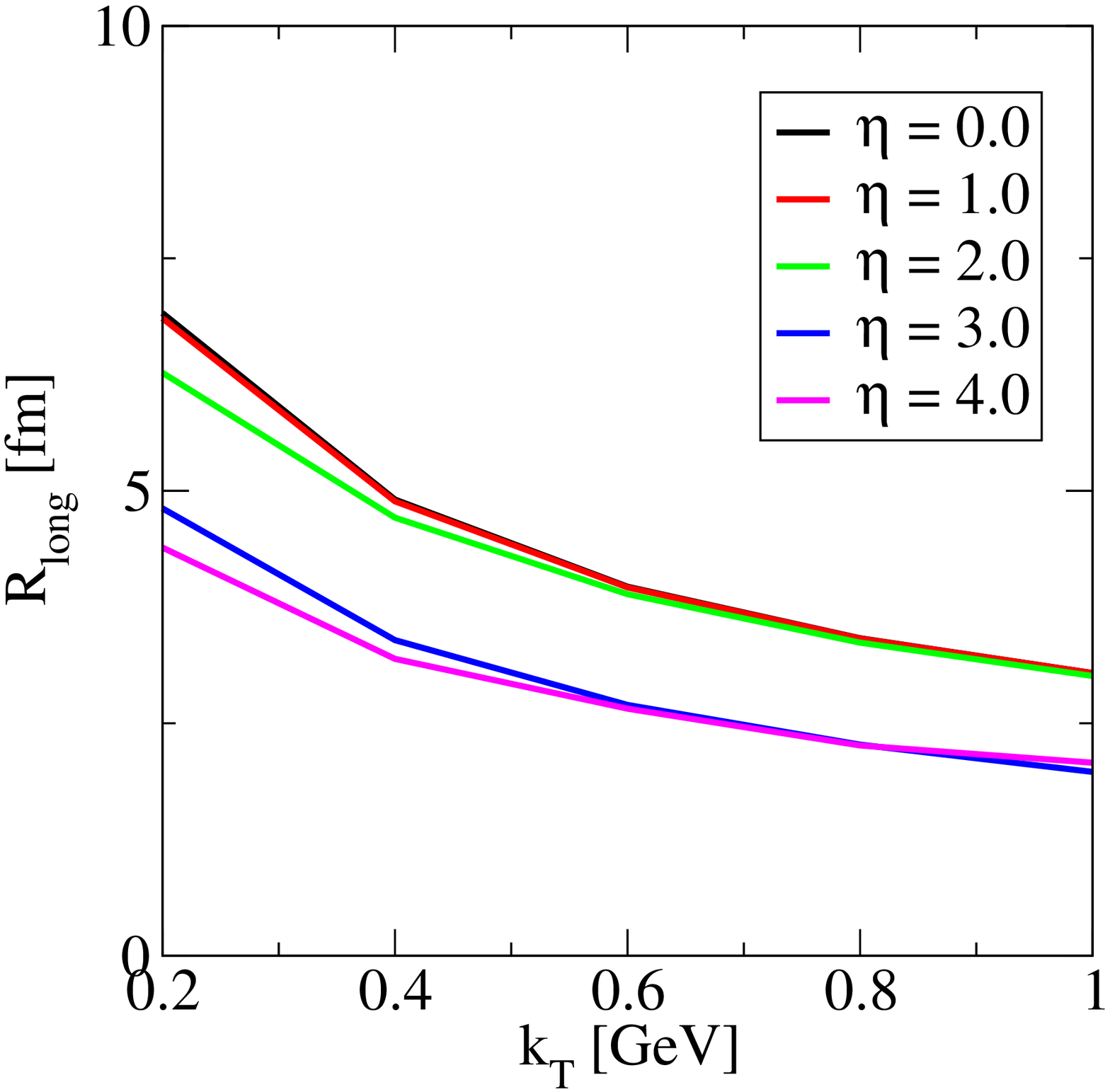}
  \caption{\label{F-long}$R_{long}$ as a function of pair momentum $k_T$ for different rapidities $\eta$ in 
an approximate boost invariant expansion (left), a longitudinally accelerating scenario with continuous particle
emission (middle) and a longitudinally accelerating scenario with sudden breakup (right).}
\end{figure}

Finally, in Fig.~\ref{F-long} I compare the three different scenarios with respect to $R_{long}$.
Due to $\eta \approx \eta_s$ in this scenario but $\eta > \eta_s$ in the presence of longitudinal acceleration, the
scaled falloff with $k_T$ is the same in all three cases (since the final $dN/d\eta$ is fixed), the correlation radius
normalization however is given by $\eta_s$ and this reduces the overall normalization. The radii in
the sudden breakup scenario are systematically larger than in the continuous emission one, reflecting the fact
that freeze-out happens on average later and hence at larger spatial extension. The sudden drop at $\eta=3$ is
probably an artefact of the model getting sensitive to the target/projectile fragmentation region which
is not well described by thermal physics.

\section{Conclusions}

The most remarkable finding is that to first order there is hardly any influence of the expansion and
freeze-out scenario on the rapidity dependence of HBT correlations. The reason for this is  that the
measured single particle spectra and $dN/d\eta$ already provide rather stringent constraints on the expansion
pattern.
The most pronounced effect of a longitudinally  
accelerated expansion is the deviation from the scaling $\eta\approx \eta_s$, ultimately
leading to a smaller $R_{long}$ in the presence of acceleration while keeping the correct falloff in $k_T$.
For the most part changes in the sideward and outward correlation radii are minor. 
Since the data practially rule out strong continuous emission throughout the lifetime by
$R_{out}/R_{side} \approx 1$, the correlation radii are dominated only by the state of the system at breakup,
not by the evolution path. It seems that for reasonable evolution scenarios
which are in agreement with other hadronic data the $\eta$ dependence of HBT radii is
primarily determined by the amount of matter at given $\eta$. Thus, photons might be a more promising
tool in order to study differences in the evolution history \cite{Photons}.


\begin{theacknowledgments}
I would like to thank B.~M\"{u}ller and J.~Ruppert for helpful discussions during the preparation of this paper.
This work was supported by the DOE grant DE-FG02-96ER40945 and a Feodor
Lynen Fellowship of the Alexander von Humboldt Foundation.
\end{theacknowledgments}



\bibliographystyle{aipproc}   

\begin{thebibliography}{99}

\bibitem{Bjorken}
J.~D.~Bjorken,
Phys.\ Rev.\ D {\bf 27} 140 (1983).

\bibitem{Landau}
L.~D.~Landau,
Izv.\ Akad.\ Nauk\  SSSR, Physics Series {\bf 17} 51 (1953).

\bibitem{Steinberg}
P.~Steinberg, nucl-ex/0405022.

\bibitem{3dh}
see e.g. T.~Hirano and Y.~Nara, J.\ Phys.\ G {\bf 31} (2005) S1, also C.~Nonaka, private communication.

\bibitem{RHIC}
  T.~Renk,
  Phys.\ Rev.\ C {\bf 70} (2004) 021903.

\bibitem{Synopsis}
  T.~Renk,
  J.\ Phys.\ G {\bf 30} (2004) 1495.

\bibitem{HBTReport}
U.~A.~Wiedemann and U.~W.~Heinz,
Phys.\ Rept.\  {\bf 319} (1999) 145.

\bibitem{PHENIX}
S.~S.~Adler et al. [PHENIX collaboration].
nucl-ex/0507004.

\bibitem{Machcones}
T.~Renk and J.~Ruppert,
hep-ph/0509036.

\bibitem{Photons}
  T.~Renk,
  Phys.\ Rev.\ C {\bf 71}, 064905 (2005).


\end{thebibliography}


\end{document}